\documentclass[aps,floats]{revtex4}
\usepackage{amsmath,amssymb}
\usepackage{graphicx,epsfig}

\begin{document}
\bibliographystyle {plain}

\def\oppropto{\mathop{\propto}} 
\def\opsimeq{\mathop{\simeq}}
\def\opoverderline{\mathop{\overline}}
\def\operarrow{\mathop{\longrightarrow}}
\def\opsim{\mathop{\sim}}

\def\fig#1#2{\includegraphics[height=#1]{#2}}
\def\figx#1#2{\includegraphics[width=#1]{#2}}


\title{ Statistics of first-passage times in disordered systems  \\
using backward master equations and their exact renormalization rules } 


 \author{ C\'ecile Monthus and Thomas Garel }
  \affiliation{ Institut de Physique Th\'{e}orique, CNRS and CEA Saclay,
 91191 Gif-sur-Yvette, France}

\begin{abstract}
We consider the non-equilibrium dynamics of disordered systems as defined by a master equation involving transition rates between configurations (detailed balance is not assumed). To compute the important dynamical time scales in finite-size systems without simulating the actual time evolution which can be extremely slow, we propose to focus on first-passage times that satisfy 'backward master equations'. Upon the iterative elimination of configurations, we obtain the exact renormalization rules that can be followed numerically. To test this approach, we study the statistics of some first-passage times for two disordered models : (i) for the random walk in a two-dimensional self-affine random potential of Hurst exponent $H$, we focus on the first exit time from a square of size $L \times L$ if one starts at the square center. (ii) for the dynamics of the ferromagnetic Sherrington-Kirkpatrick model of $N$ spins, we consider the first passage time $t_f$ to zero-magnetization  when starting from a fully magnetized configuration. Besides the expected linear growth of the averaged barrier $\overline{\ln t_{f}} \sim N$, we find that the rescaled distribution of the barrier $(\ln t_{f})$ decays as $e^{- u^{\eta}}$ for large $u$ with a tail exponent of order $\eta \simeq 1.72$. This value can be simply interpreted in terms of rare events if the sample-to-sample fluctuation exponent for the barrier is $\psi_{width}=1/3$.

\end{abstract}

\maketitle

\section{ Introduction }

In statistical physics, any large-scale universal behavior is expected to 
come from some underlying renormalization ('RG') procedure that eliminates
all the details of microscopic models.
 For the non-equilibrium dynamics of disordered systems, we have recently proposed
a strong disorder renormalization procedure
 in configuration space that can be defined for
any master equation \cite{rgmaster,rgmastereq,rgu2d} :
it is based on the iterative elimination of the smallest barrier
 remaining in the system, and thus generalizes the real-space strong disorder
procedures that had been previously defined for random walks in
one-dimensional random media
\cite{sinairg,sinaibiasdirectedtraprg,readiffrg,rfimrg,traprg}.
However, as for all strong disorder renormalization procedures (see \cite{review} for a review), the results are asymptotically exact only near ``Infinite disorder fixed points'' :
for the dynamical problems defined by a master equation, this means
that the strong disorder renormalization procedure will give asymptotically exact results
only if the renormalized distribution of barriers becomes broader and broader upon
iteration (see  \cite{rgmaster} for a more detailed discussion).
In the present paper, we show that one can obtain exact renormalization rules,
{\it without any strong disorder hypothesis,} if one considers the 'backwards master
equation' satisfied by first-passage times.
It turns out that the renormalization rules for the transition rates are
formally identical to the strong disorder rules introduced in  \cite{rgmaster,rgmastereq},
but the interpretation, the goals, and the validity of the two approaches
are different, as we explain in more details below.

From a numerical point of view, the main limitation of Monte-Carlo dynamical simulations 
of disordered systems
is that the dynamics in the presence of quenched disorder
 becomes extremely slow as the system size increases (see for instance
the introduction of our recent work \cite{conjugate} and references therein).
It is thus important to develop other methods to characterize the dynamical properties
of disordered systems { \it without simulating the dynamics}.
For instance in our previous work \cite{conjugate}, 
we have proposed to use the mapping between
any master equation satisfying detailed balance and some Schr\"odinger equation
in configuration space, to obtain the largest relaxation time of the dynamics
via any eigenvalue method able to compute the energy of the first excited state
of the associated quantum Hamiltonian. Here we propose another strategy
based on the 'backwards master equation' satisfied by first-passage times.
The fact that first-passage times satisfy 'backwards master equation'
 is of course very well-known and can be found
in most textbooks on stochastic processes (see for instance 
\cite{gardiner,vankampen,risken,redner}). 
In the field of disordered systems, the backward Fokker-Planck equation
has been very much used to characterize
the dynamics of a single particle in a random medium (see for instance \cite{Comtet_Dean,Dean_Maj,Maj_Comtet,Maj_Comtet2,sinaibiasdirectedtraprg,Maj_Comtet3}),
but to the best of our knowledge, this approach has not yet been used in higher dimension,
 nor for many-body problems.
To test the present approach, we compute the statistics of
first-passage times over the disordered samples of a given size
for two disordered models (i) a random walk in a two-dimensional random potential
(ii)  a mean-field spin model.

The paper is organized as follows.
In section \ref{secBack}, we recall that
first-passage times satisfy 'backward master equation'.
In section \ref{secRGrules}, we derive the corresponding renormalization rules 
and discuss the similarities and differences with respect to strong disorder
renormalization procedures.
We then apply this approach to two types of disordered models : 
section \ref{secu2d} concerns 
 the problem of a random walk in a two-dimensional self-affine potential, 
and  section \ref{secSKferro} is devoted to the dynamics of  
the ferromagnetic Sherrington-Kirkpatrick model.
Our conclusions are summarized in section 
\ref{secConclusion}.

\section{ Reminder on first-passage times and backward master equations }

\label{secBack}

\subsection{ Master equation defining the stochastic dynamics }

In statistical physics, it is convenient to consider
continuous-time stochastic dynamics defined by a 'forward' master equation of the form
\begin{eqnarray}
\frac{ dP_t \left({\cal C} \right) }{dt}
= \sum_{\cal C '} P_t \left({\cal C}' \right) 
W \left({\cal C}' \to  {\cal C}  \right) 
 -  P_t \left({\cal C} \right) W_{out} \left( {\cal C} \right)
\label{master}
\end{eqnarray}
that describes the evolution of the
probability $P_t ({\cal C} ) $ to be in  configuration ${\cal C}$
 at time t.
The notation $ W \left({\cal C}' \to  {\cal C}  \right) $ 
represents the transition rate per unit time from configuration 
${\cal C}'$ to ${\cal C}$, and 
\begin{eqnarray}
W_{out} \left( {\cal C} \right)  \equiv
 \sum_{ {\cal C} '} W \left({\cal C} \to  {\cal C}' \right) 
\label{wcout}
\end{eqnarray}
represents the total exit rate out of configuration ${\cal C}$.

\subsection{ Backward Master Equation satisfied by the first-passage time }

Let us now focus on the following problem : suppose the dynamics starts
at $t=0$ in configuration ${\cal C}$, and one is interested in the 
random time $t$ where the dynamics will 
reach for the first time any configuration belonging to
 a given set $A$ of 'target' configurations.
As is well known 
(see for instance the textbooks \cite{gardiner,vankampen,risken,redner}),
the mean first-passage time $\tau^{(A)}({\cal C})=<t>$ 
(where the notation $<.>$ represents the average
 with respect to the dynamical trajectories) satisfies the following
'backward master equation' for all configurations ${\cal C}$ not in the set $A$
\begin{eqnarray}
\sum_{\cal C \ '} 
W \left({\cal C}\  \to  {\cal C}'  \right) \tau^{(A)} \left({\cal C}\ ' \right) 
 -  W_{out} \left( {\cal C} \right)  \tau^{(A)} \left({\cal C} \right) = -1
\label{backwardT}
\end{eqnarray}
whereas all configurations in the set $A$ satisfy the
 boundary conditions 
\begin{eqnarray}
\tau^{(A)}({\cal C} \in A)=0
\label{bcT}
\end{eqnarray}

The derivation of Eq. \ref{backwardT} consists in considering
what happens during the first time interval $[0, dt]$ 
if the system is in configuration ${\cal C}$ at $t=0$ : at time $dt$,
the system is either in configuration ${\cal C}'$ with probability 
$[ W \left({\cal C}  \to  {\cal C}'  \right) dt]$, 
in which case the remaining mean time is $ \tau^{(A)} ({\cal C}\ ' )$, or the system is still 
in configuration  ${\cal C}$ with probability 
$[1- W_{out} \left( {\cal C} \right) dt]$, 
in which case the remaining mean-time is $ \tau^{(A)} ({\cal C}\  )$.
 By consistency, the mean
first passage time has thus to satisfy at first order in $dt$
\begin{eqnarray}
\tau^{(A)}({\cal C}) =  dt + \sum_{\cal C \ '} 
\left[ W \left({\cal C}\  \to  {\cal C}'  \right) dt \right] \ 
 \tau^{(A)} ({\cal C}\ ' ) 
+ \left[1- W_{out} \left( {\cal C} \right) dt \right] \tau^{(A)}({\cal C}) 
\label{derivation}
\end{eqnarray}
yielding Eq. \ref{backwardT}.

The backward master equations of Eq. \ref{backwardT}
can be solved numerically by any method appropriate for
 linear equations with fixed right hand-side. 
In the next section, we show that they satisfy exact renormalization rules.

\section{ Renormalization rules for first-passage time properties }

\label{secRGrules}

\subsection{ Iterative elimination of configurations }

If one eliminates iteratively the configurations from the system
of Eqs \ref{backwardT} satisfied by the first-passage times,
the renormalized equations for the surviving configurations
keep the same form, but with renormalized transition rates $W^R$ and renormalized
right-hand sides $K^R$
\begin{eqnarray}
\sum_{\cal C \ '} 
W^R \left({\cal C}\  \to  {\cal C}'  \right) \tau^{(A)} \left({\cal C}\ ' \right) 
 -  W_{out}^R \left( {\cal C} \right)  \tau^{(A)} \left({\cal C} \right) 
= - K^R \left({\cal C} \right) 
\label{rgbackwardT}
\end{eqnarray}

This equation for ${\cal C}={\cal C}_0$ can be used to eliminate 
$\tau^{(A)} \left({\cal C}_0 \right)$ via
\begin{eqnarray}
\tau^{(A)} \left({\cal C}_0 \right) =\frac{1}{W_{out}^{R} \left( {\cal C}_0 \right)}
\left[ \sum_{\cal C \ ''} 
W^{R} \left({\cal C}_0\  \to  {\cal C}''  \right) \tau^{(A)} \left({\cal C}\ '' \right) 
+ K^{R}\left({\cal C}_0 \right) \right]
\label{elimTr0}
\end{eqnarray}
Upon the elimination of the configuration ${\cal C}_0$, 
the renormalized coefficients $W^R$ and $K^R$ evolve 
according to the following renormalization rules for the surviving configurations 
${\cal C}$
\begin{eqnarray}
W^{Rnew} \left({\cal C}\  \to  {\cal C}'  \right) && =  
W^R \left({\cal C}\  \to  {\cal C}'  \right)+ \frac{W^R \left({\cal C}\  \to  {\cal C}_0  \right) W^R \left({\cal C}_0\  \to  {\cal C}'  \right)}{W_{out}^R \left( {\cal C}_0 \right)} \nonumber \\
W_{out}^{Rnew} \left( {\cal C} \right) && =W_{out}^R \left( {\cal C} \right)
- \frac{W^R \left({\cal C}\  \to  {\cal C}_0  \right) W^R \left({\cal C}_0\  \to  {\cal C}  \right)}{W_{out}^R \left( {\cal C}_0 \right)} \nonumber \\
K^{Rnew}\left({\cal C}\right) && = K^R\left({\cal C} \right)
+ \frac{W^R \left({\cal C}\  \to  {\cal C}_0  \right)}{W_{out}^R \left( {\cal C}_0 \right)} K^R\left({\cal C}_0 \right)
 \label{rgrulesT}
\end{eqnarray}

\subsection{Renormalization rules for other observables satisfying 'backward master equation' }

Since other observables are known to satisfy similar 'backward master equations',
it is interesting to discuss here their renormalization rules
and to compare with Eqs \ref{rgrulesT}.

\subsubsection { Higher moments of first-passage times }

Above we have considered the first moment $\tau^{(A)}({\cal C})=<t>$
of the first-passage time in the set $A$ when starting in configuration ${\cal C}$.
However, one may consider the higher moments $\tau^{(A)}_n({\cal C})=<t^n>$
that satisfy the following
'backward master equation' \cite{gardiner,vankampen,risken,redner})
for all configurations ${\cal C}$ not in the set $A$
\begin{eqnarray}
\sum_{\cal C \ '} 
W \left({\cal C}\  \to  {\cal C}'  \right) \tau_n \left({\cal C}\ ' \right) 
 -  W_{out} \left( {\cal C} \right)  \tau_n \left({\cal C} \right) = - n \tau_{n-1} \left({\cal C} \right)
\label{backwardTn}
\end{eqnarray}
whereas all configurations in the set $A$ satisfy the
 boundary conditions 
\begin{eqnarray}
\tau_n({\cal C} \in A)=0
\label{bcTn}
\end{eqnarray}
The derivation of Eq. \ref{backwardTn} consists again in considering
what happens during the first time interval $[0, dt]$ (see explanations before 
Eq \ref{derivation}).
The higher moments of first-passage times
 can be thus computed one after the other : if one knows the moments of order $(n-1)$,
one can compute the moments of order $n$ via the same renormalization rules of Eq. 
\ref{rgrulesT} : the only change will be in the initial condition for the right handside
that will read $K^{initial}_n\left({\cal C}\right)= n \tau_{n-1} \left({\cal C} \right)$
instead of $K^{initial}_{n=1}\left({\cal C}\right)=1$.

\subsubsection { Escape probabilities }

The simplest quantities that satisfy some backward master equation
are the escape probabilities.
Suppose the dynamics starts
at $t=0$ in configuration ${\cal C}$, and one is interested into the 
probability $E_{B/A}({\cal C})$ to reach first any configuration belonging to
 a set $B$ of configurations before any configuration belonging to
another set $A$ of configurations.
As is well known (see for instance the textbooks \cite{gardiner,vankampen,risken,redner}),
this escape probability $E_{B/A}({\cal C})$ satisfies the following
'backward master equation' for all configurations ${\cal C}$ neither in the set
 $A$ nor in the set $B$
\begin{eqnarray}
\sum_{\cal C \ '} 
W \left({\cal C}\  \to  {\cal C}'  \right) E_{B/A} \left({\cal C}\ ' \right) 
 -  W_{out} \left( {\cal C} \right)  E_{B/A} \left({\cal C} \right) = 0
\label{backwardE}
\end{eqnarray}
whereas the configurations in the set $A$ or in the set $B$ satisfy the
 boundary conditions 
\begin{eqnarray}
E_{B/A}({\cal C} \in A)=0 \\
E_{B/A}({\cal C} \in B)=1 \\
\label{bcE}
\end{eqnarray}

The backward master Eq. \ref{backwardE} does not contain any right handside
in contrast to Eq. \ref{backwardT} :
the iterative elimination of configurations will lead
to renormalized transition rates that follows the same two first rules of Eq. 
\ref{rgrulesT}.

\subsection{ Similarities and differences with the strong disorder
 renormalization of Refs \cite{rgmaster,rgmastereq}   }

It turns out that the renormalization rules for the transition rates given
in the two first lines of Eq. \ref{backwardT}  are
formally identical to the strong disorder rules 
introduced in  \cite{rgmaster,rgmastereq}. It is thus important to stress here
why the interpretation, the goals, and the validity of the two approaches
are significantly different :

(i) The present renormalization rules are exact
 for any dynamics defined by a master equation.
But they yield results only for observables
like first-passage times that satisfy backwards master equations with fixed 
right hand-side.

(ii) On the contrary, the strong disorder renormalization procedure 
introduced in  \cite{rgmaster,rgmastereq} aims to renormalize the forward
master equation of Eq. \ref{master}, i.e. the full time evolution of the probability
distribution $P_t({\cal C})$.
It will become asymptotically exact at large times 
only for dynamics governed by an 'infinite disorder
fixed point' (see more details in \cite{rgmaster}).
However whenever it is the case, it can yield results
for any universal observable (i.e. exponents or rescaled distributions).

\section{Random walk in a two-dimensional self-affine potential  }

\label{secu2d}

\begin{figure}[htbp]
 \includegraphics[height=6cm]{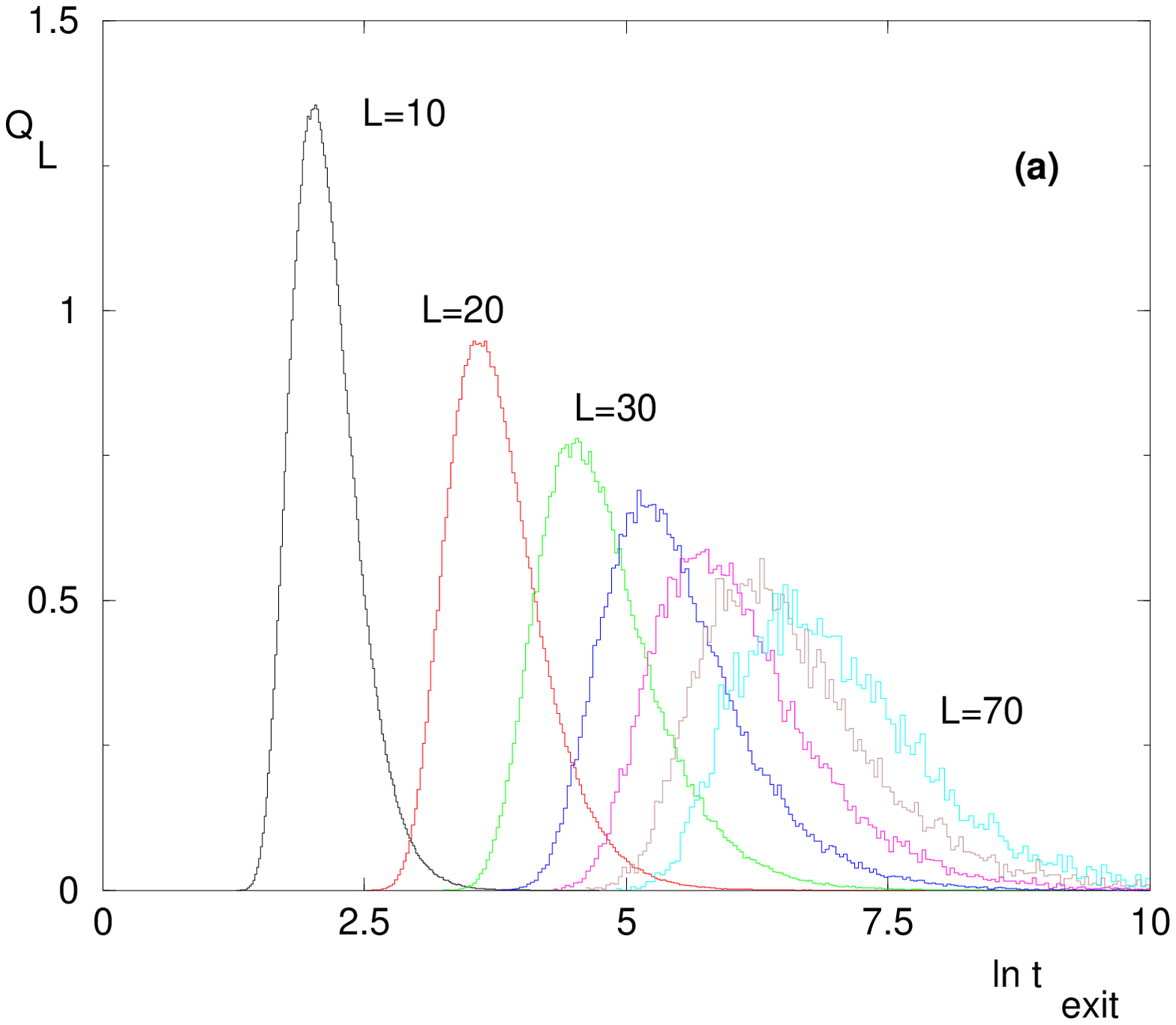}
\hspace{2cm}
 \includegraphics[height=6cm]{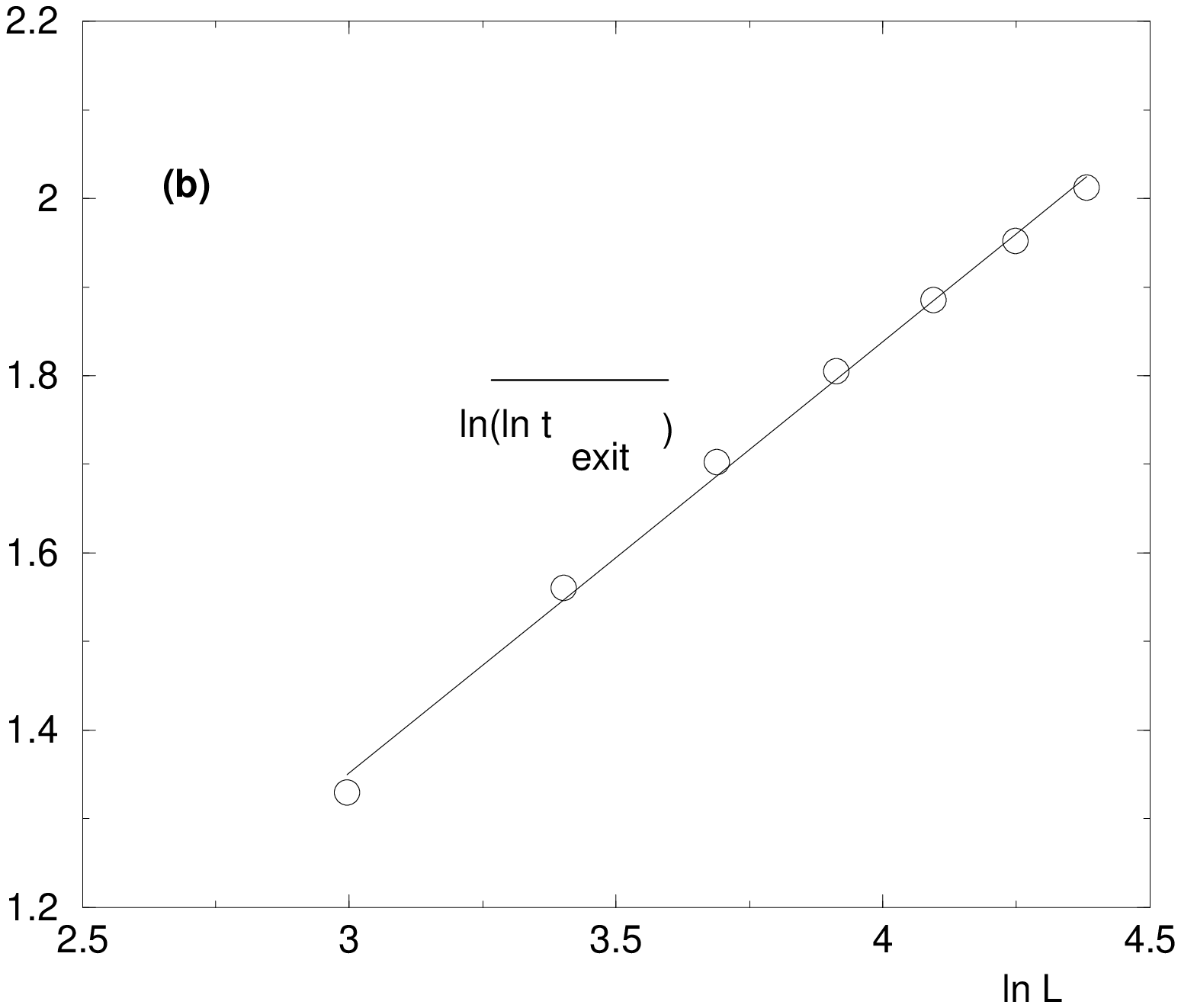}
\vspace{1cm}
\caption{ (Color on line) 
 Statistics of the first exit time $t_{exit}$ from a square of size $L \times L$
when starting at the center 
 for the random walk in a self-affine random potential of Hurst exponent
 $H=0.5$ :
(a) Probability distribution 
 $Q_{L}(\Gamma_{exit}=\ln t_{exit})$ for $L=20,30,40,50,60,70$ ;
(b) the log-log plot of the disorder-average 
$ \overline{\Gamma_{exit}}(L)=  \overline{\ln t_{exit} }(L) $ 
corresponds to the barrier exponent $\psi=H=0.5$ (Eq. \ref{psiu2d}). }
\label{figu2d}
\end{figure}

In this section, we apply the method of the previous section
to the continuous-time random walk of a particle
 in a two-dimensional self-affine quenched random potential of Hurst exponent $H=0.5$.
Since we have studied recently in \cite{rgu2d} the very same model via some strong disorder
renormalization procedure, we refer the reader to \cite{rgu2d} and references therein
for a detailed presentation of the model and of the numerical method to generate the random potential. Here we simply recall what is necessary for the present approach.

We consider a two-dimensional square lattice of size $L\times L$.
 The continuous-time random walk in the random potential $U(\vec r)$
is defined by the master equation
\begin{eqnarray}
\frac{ dP_t \left({\vec r} \right) }{dt}
= \sum_{\vec r \ '} P_t \left({\vec r}\ ' \right) 
W \left({\vec r}\ ' \to  {\vec r}  \right) 
 -  P_t \left({\vec r} \right) W_{out} \left( {\vec r} \right)
\label{masteru2d}
\end{eqnarray}
where the transition rates are given by the Metropolis choice
at temperature $T$ (the numerical data presented below correspond to $T=1$)
\begin{eqnarray}
W \left( \vec r \to \vec  r \ '  \right)
= \delta_{<\vec r, \vec r\ ' >} 
\  {\rm min} \left(1, e^{-  (U(\vec r \ ' )-U(\vec r ))/T } \right)
\label{metropolisu2d}
\end{eqnarray}
where the factor $\delta_{<\vec r, \vec r\ ' >}$
 means that the two positions
are neighbors on the two-dimensional lattice.
The random potential $U(\vec r)$ is self-affine with Hurst exponent $H=0.5$
\begin{eqnarray}
\overline{ \left[ U(\vec r) -U(\vec r \ ') \right]^2 }
\opsimeq_{ \vert \vec r - \vec r \ ' \vert \to \infty}
 \vert \vec r - \vec r \ ' \vert^{2H}
\label{correU2d}
\end{eqnarray}

We focus here on the first-passage time $\tau^{(A)}({\cal C}_0)$
corresponding to the following conditions :
(i) the initial configuration
${\cal C}_0$ is the center of the square $(x_0=L/2,y_0=L/2)$
(ii) the set $A$ of 'target configurations' 
is the set of all boundary sites of
the square, i.e. having $x=1$, $x=L$, $y=1$ or $y=L$.
The first-passage time $\tau^{(A)}({\cal C}_0)$ thus corresponds here to
the first exit time $t_{exit}$ from the square $L \times L$ when starting
at the center. The appropriate variable is actually the barrier defined as
\begin{eqnarray}
\Gamma_{exit} \equiv \ln t_{exit}
\label{defbarrier}
\end{eqnarray}

On Fig. \ref{figu2d} (a), we show
 the corresponding probability distribution 
$Q_L(\Gamma_{exit}\equiv \ln t_{exit})$
for various sizes  $20 \leq L \leq 80$ with a statistics 
of  $9.10^5 \geq n_s(L) \geq 36.10^2$  disordered samples. 

As shown by the log-log plot of Fig. \ref{figu2d} (b),
we find that the disorder-averaged value ${\overline \Gamma_{exit}(L)}$
scales as
\begin{eqnarray}
 \overline{\Gamma_{exit}}(L)  \oppropto_{L \to \infty}  L^{\psi} 
\label{psiscaling}
\end{eqnarray}
with a barrier exponent $\psi$ of order
\begin{eqnarray}
\psi=H=0.5
\label{psiu2d}
\end{eqnarray}
These results are in agreement with 
scaling arguments on barriers \cite{Mar83,jpbreview},
 with the strong disorder renormalization approach of \cite{rgu2d},
and with the computation of the relaxation time to equilibrium \cite{conjugate}.

\section{Dynamics of ferromagnetic Sherrington-Kirkpatrick model  }

\label{secSKferro}

\begin{figure}[htbp]
 \includegraphics[height=6cm]{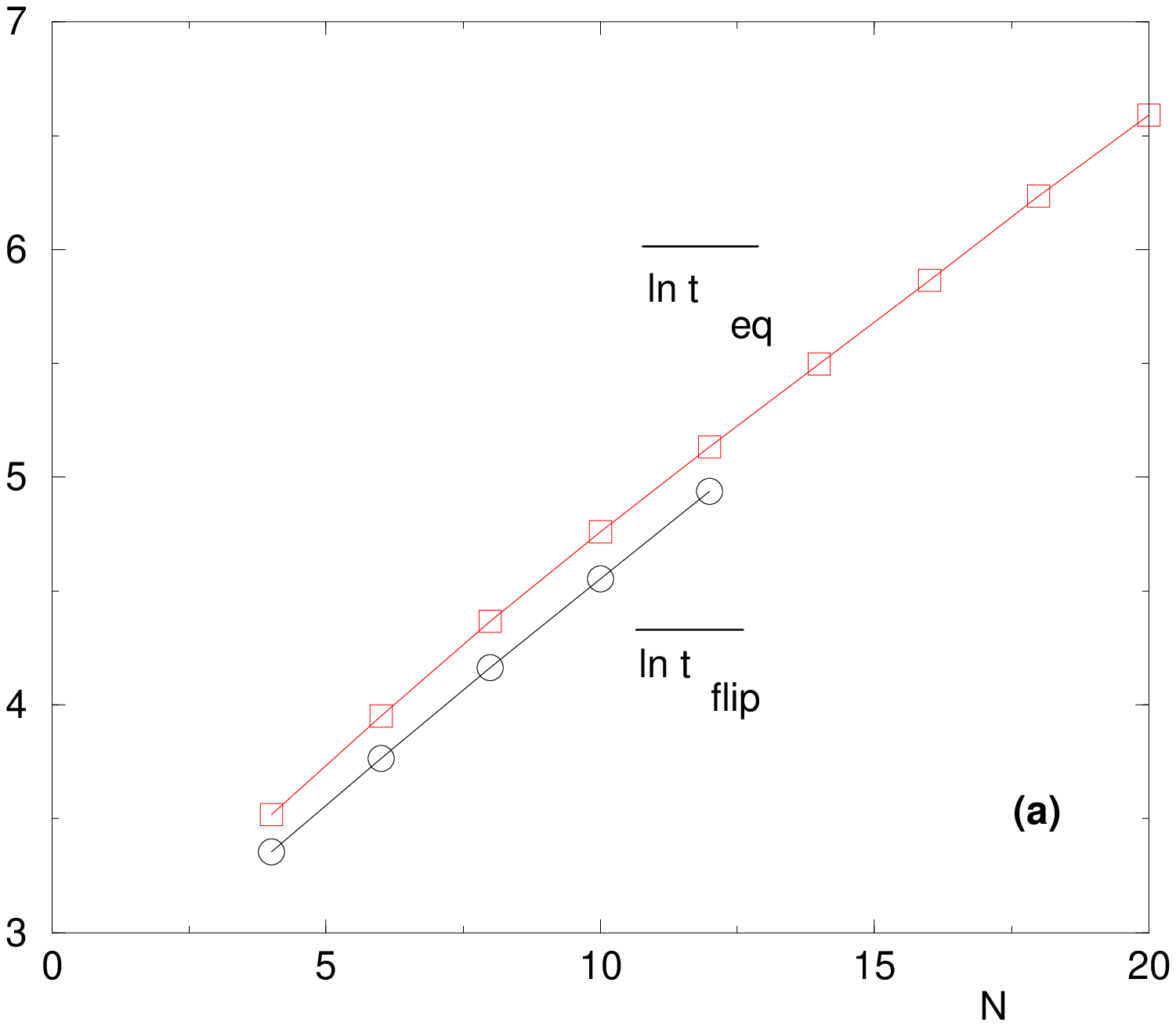}
\hspace{2cm}
 \includegraphics[height=6cm]{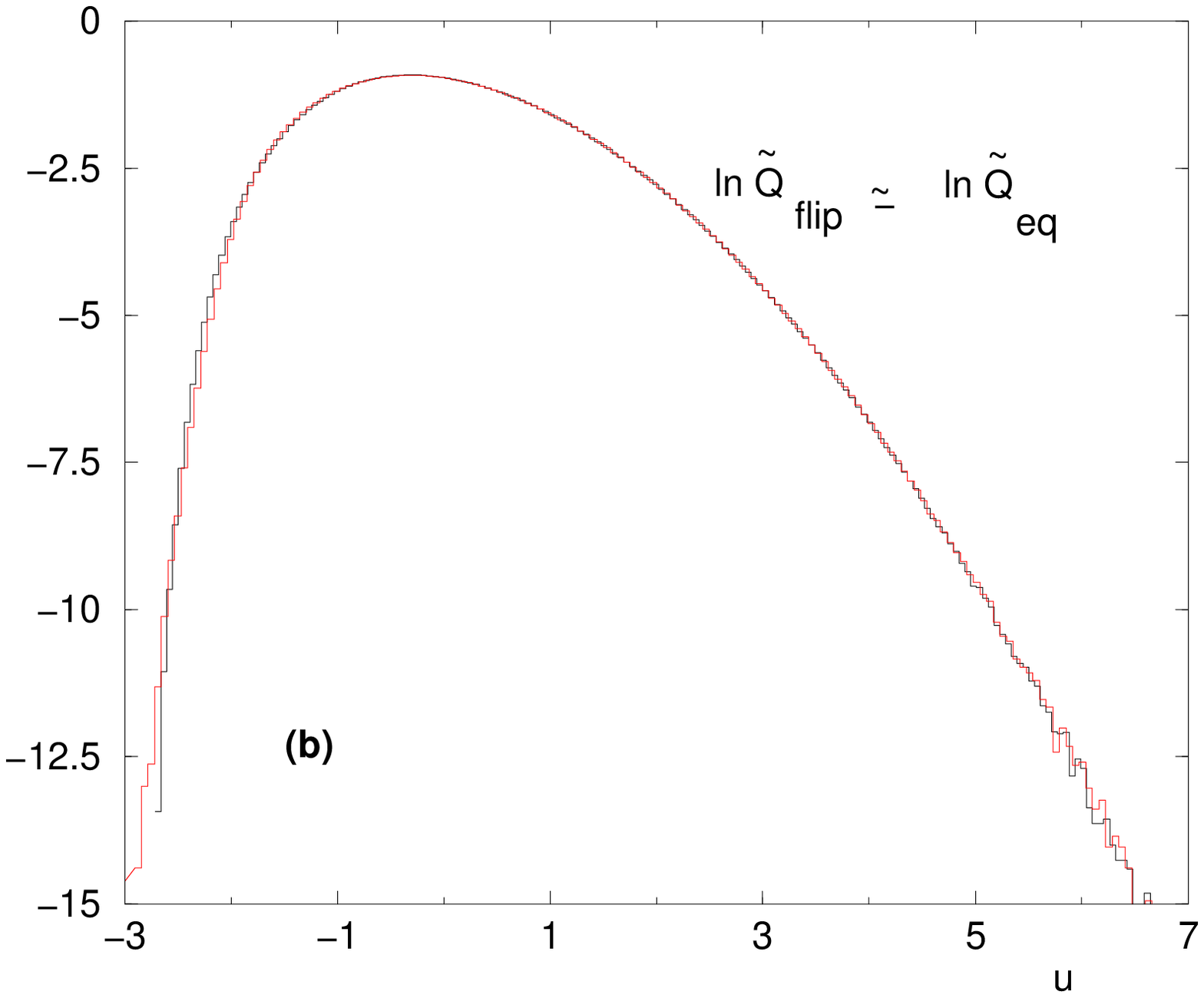}
\vspace{1cm}
\caption{ (Color on line) 
 Statistics of the first time $t_{flip}$ where the magnetization vanishes,
 for the ferromagnetic Sherrington-Kirkpatrick model of $N$ spins 
(Eq \ref{jijrandomferro}) :
(a) the disorder-average $ \overline{\ln t_{flip}}(N) $ grows linearly
with $N$ (Eq. \ref{ferroav}). The disorder-average 
$ \overline{\ln t_{eq}}(N) $ associated to the largest relaxation time
$t_{eq}(N) $ towards equilibrium
as computed from the method of Ref \cite{conjugate} is also shown
for comparison.
(b) The rescaled probability distribution 
 ${\tilde Q}_{flip}(u)$ of Eq. \ref{defQrescaled}, shown here
in log scale to see the tail of Eq. \ref{defeta}, exactly coincides with
the rescaled probability distribution 
 ${\tilde Q}_{eq}(u)$ as computed from the method of Ref \cite{conjugate} : 
the tail exponent is for both of order $\eta \simeq 1.72$ 
(Eq. \ref{etaferrorandom} ).  }
\label{figSKferro}
\end{figure}

As an example of application to a many-body disordered system,
we consider in this section the ferromagnetic
Sherrington-Kirkpatrick model  where a configuration 
${\cal C}=\{S_i\}$ of $N$ spins $S_i =\pm 1$ has for energy 
\begin{eqnarray}
 U = - \sum_{1 \leq i <j \leq N} J_{ij} S_i S_j
\label{defSK}
\end{eqnarray}
where the coupling $J_{ij}$ between two spins $S_i$ and $S_j$ contains
a non-random ferromagnetic part $J_0$ and a random Gaussian 
part ${\tilde J}_{ij}$ of
zero-mean $\overline{ {\tilde J}_{ij}}=0$ and variance unity 
$\overline{ {\tilde J}_{ij}^2}=1$ with the appropriate mean-field rescalings
\cite{SKmodel,almeida,toulouse,nishimori}
\begin{eqnarray}
 J_{ij} = \frac{J_0}{N-1} + \frac{{\tilde J}_{ij}}{\sqrt{N-1}}
\label{jijrandomferro}
\end{eqnarray}
Here we consider the values $J_0=2$ and temperature $T=1$ 
where the model is in its ferromagnetic phase 
\cite{SKmodel,almeida,toulouse,nishimori}
to study its dynamical properties.
The Metropolis dynamics corresponds to the master equation of Eq. \ref{master}
in configuration space with the transition rates
\begin{eqnarray}
W \left( {\cal C} \to  {\cal C} '  \right)
= \delta_{<{\cal C}, {\cal C}\ ' >} 
\  {\rm min} \left(1, e^{-  (U({\cal C} \ ' )-U({\cal C} ))/T } \right)
\label{metropolisSK}
\end{eqnarray}
where the factor $\delta_{<{\cal C}, {\cal C}\ ' >}$ 
means that the two configurations are related by a single spin flip.

We focus here on the first-passage time $\tau^{(A)}({\cal C}_0)$
corresponding to the following conditions :
(i) the initial configuration
${\cal C}_0$ is the fully ferromagnetic configuration 
of magnetization $ M_N=\sum_{i=1}^N S_i=N$ where all spins are $S_i=+1$.
(ii) the set $A$ of 'target configurations' 
is the set of all configurations of zero magnetization $M_N=\sum_{i=1}^N S_i=0$
(we consider only even N).
The first-passage time $\tau^{(A)}({\cal C}_0)$ thus corresponds here to
the first time $t_{flip}$ where the magnetization $M_N$ vanishes.

We have computed the distribution $Q_L$ of the barrier defined as
\begin{eqnarray}
\Gamma_{flip} \equiv \ln t_{flip}
\label{defbarrierflip}
\end{eqnarray}
 over the disordered samples
of even sizes $4 \leq N \leq 12$
with a statistics 
of  $2.10^8 \geq n_s(L) \geq 6.10^2$  samples. 
As a comparison, we have also computed the distribution of the barrier
 $\Gamma_{eq} \equiv \ln t_{eq}$, where $t_{eq}$ is defined as 
the largest relaxation time towards equilibrium via the method described
in our previous work \cite{conjugate}. 
Since the system is in its ferromagnetic phase, one expects that
the disorder-average of the barrier grows as
\begin{eqnarray}
 \overline{\Gamma_{flip}}(N) = \overline{\ln t_{flip}} \oppropto_{N \to \infty}  N 
\label{ferroav}
\end{eqnarray}
and this is indeed what we measure both for $\overline{\Gamma_{flip}}(N)$
and for $\overline{\Gamma_{eq}}(N)$ as shown on Fig. \ref{figSKferro} (a).
The width $\Delta(N)$ of the barrier distribution is expected
 to grow with a subleading exponent $0<\psi_{width}<1$
\begin{eqnarray}
\Delta(N) \equiv
 \left( \overline{\Gamma_{flip}^2}(N)
 - (\overline{\Gamma_{flip}}(N))^2\right)^{1/2} 
  \oppropto_{N \to \infty} N^{\psi_{width}}
\label{ferrowidth}
\end{eqnarray}
but we are not aware of any theoretical prediction or any previous numerical measure
of this sample-to-sample fluctuation exponent $\psi_{width}$. This is in contrast with
the spin-glass Sherrington-Kirkpatrick model
corresponding to $J_0=0$, where the barrier exponent has been much studied
either theoretically \cite{rodgers,horner}
or numerically 
\cite{young,vertechi,colbourne,billoire,janke,conjugate}.

With our numerical data limited to small sizes $4 \leq N \leq 12$,
we see already the expected linear behavior of the disorder-average of Eq.
\ref{ferroav} as shown on Fig. \ref{figSKferro} (a), but we are unfortunately not able to measure
the exponent $\psi_{width}$ of Eq. \ref{ferrowidth} from the variance.
However, since for these small sizes we can study a large statistics of disordered samples, we have measured the rescaled distribution ${\tilde Q}$ defined as
\begin{eqnarray}
Q_{L}(\Gamma_{flip}) \sim  
  \frac{1}{\Delta(N) } {\tilde Q}_{flip} 
\left( u \equiv \frac{\Gamma_{flip} - \overline{\Gamma_{flip}}(N) }{\Delta(N) }
 \right)
\label{defQrescaled}
\end{eqnarray}
We find that the rescaled distribution 
${\tilde Q} (u) $ shown on Fig. \ref{figSKferro} (b)  presents at large argument
the exponential decay
\begin{eqnarray}
\ln {\tilde Q}_{flip} (u) \oppropto_{u \to + \infty} - u^{\eta}
\label{defeta}
\end{eqnarray}
with a tail exponent of order
\begin{eqnarray}
\eta \simeq 1.72
\label{etaferrorandom}
\end{eqnarray}
We have moreover checked that the rescaled distribution 
${\tilde Q}_{flip} (u) $ exactly coincides with
the rescaled probability distribution 
 ${\tilde Q}_{eq}(u)$ as computed from the method of Ref \cite{conjugate}.

To interpret the value of Eq. \ref{etaferrorandom}, 
one may propose the following rare-event argument.
Since the system is in its ferromagnetic phase, it seems natural to expect
that the anomalously large barriers in the dynamics will correspond to
the samples that have anomalously strong ferromagnetic contributions coming from the random parts of the couplings in Eq. \ref{jijrandomferro} : with an exponentially rare probability of order $e^{- (cst) N^2}$, the $N^2$ random variables 
$ {\tilde J}_{ij}$ will be all positive. Then instead of being finite, the 
local field $h_i=\sum_j J_{ij} S_j $ on spin $S_i$ will be of order $N^{1/2}$,
and one thus expects a barrier of order $N^{3/2}$. If one plugs these values
in Eqs \ref{defQrescaled} and \ref{defeta}, one obtains, for the powers of $N$ in the 
exponentials, the consistency equation
\begin{eqnarray}
\left( \frac{3}{2}- \psi_{width} \right) \eta = 2
\label{rareferrorandom}
\end{eqnarray}
For instance $ \psi_{width}=1/2$ would correspond to $\eta=2$.
The value 
\begin{eqnarray}
 \psi_{width} = \frac{1}{3} 
\label{onethird}
\end{eqnarray}
would correspond to the tail exponent value
\begin{eqnarray}
\eta( \psi_{width}= \frac{1}{3}) = \frac{12}{7} =1.714...
\label{etarare}
\end{eqnarray}
which is extremely close to the value that we measure numerically 
(Eq. \ref{etaferrorandom}).
A tentative conclusion would thus be the following :
at the small sizes that we can study, we cannot measure the width exponent
$ \psi_{width}$ from the variance, but we can measure the tail exponent $\eta$ that contains the information on $\psi_{width}$ if one can properly identify
 the rare events that dominate the tail.
In the ferromagnetic phase considered here, we believe that 
the rare events dominating the tail are the anomalously strong ferromagnetic samples described above, so that our measure of the tail exponent of Eq. 
\ref{etaferrorandom} would point towards the value of Eq. \ref{onethird}
for the width exponent.
Of course, this type of indirect reasoning based on rare events
remains rather speculative, and a direct measure of $\psi_{width}$
from the variance for large sizes $N$ via Monte-Carlo simulations
would be very welcome (to the best of our knowledge,
the variance has only been measured up to now 
for the case $J_0=0$ in \cite{janke}).

\section{Conclusion }

\label{secConclusion}

To avoid the simulation of the dynamics of disordered systems
 which can be extremely slow, 
 we have proposed in this paper to focus on first-passage times 
that satisfy 'backward master equations'. We have shown that
these equations satisfy exact renormalization rules upon the 
 iterative elimination of configurations. We have explained the similarities
and differences with the strong disorder
 renormalization of Refs \cite{rgmaster,rgmastereq}.
 We have
 then tested numerically this approach for two types of disordered models :
 (i) for the random walk in a two-dimensional self-affine random
 potential of Hurst exponent $H=1/2$, 
we have computed the statistics of
 the first exit time from a square of size $L \times L$
if one starts at the square center.
 (ii) for the dynamics of the ferromagnetic Sherrington-Kirkpatrick model,
 we have studied the statistics of
 the first passage time $t_f$ to zero-magnetization  
when starting from a fully magnetized configuration.
We have compared with the results concerning the largest relaxation time
towards equilibrium obtained with the method of \cite{conjugate}.
Our conclusion is that the first-passage method is reliable
to measure dynamical properties of disordered systems. 
Although in some cases, it takes more CPU time than the method
of  \cite{conjugate}, it can have several advantages in other cases :

(i) it does not require the detailed balance condition (in contrast to \cite{conjugate})

(ii) the CPU time depends only
on the size of configuration space, but not at all on the disorder realization
and on the time scales involved that can be arbitrarily large. 
(in contrast to \cite{conjugate} where the convergence of the iteration method
depends on the disorder sample and on the temperature).

(iii)  the freedom in the choice of the initial condition
and of the 'target configurations',
can be useful 
to study the time scales associated to various dynamical processes
(whereas the  method of \cite{conjugate} focuses
on the largest relaxation time towards equilibrium).

\section*{Acknowledgements }

It is a pleasure to thank A. Billoire, J.P. Bouchaud, A. Bray and M. Moore
for discussion or correspondence on the statistics
 of dynamical barriers in mean-field spin-glasses.

\end{document}